\documentclass[a4paper]{paper} 
\usepackage[margin=2.5cm]{geometry}
\usepackage{lipsum}
\usepackage{xcolor}
\sectionfont{\large\sf\bfseries\color{black!70!blue}} 
\usepackage{graphicx} 
\usepackage{authblk}

\usepackage{braket}
\usepackage{physics}
\usepackage{multirow}
\usepackage{siunitx}

\usepackage{subcaption}
\usepackage{algorithm}
\usepackage[algo2e]{algorithm2e} 
\usepackage{hyperref}
\usepackage{cleveref}
\newcommand{\figref}[1]{Figure {\ref{#1}}}
\newcommand{\secref}[1]{Section {\ref{#1}}}

\renewcommand{\eqref}[1]{Eq. ({\ref{#1}})}
\usepackage{biblatex}
\bibliography{Ref.bib}

\begin{document}
\title{\centering A supplemental investigation of non-linearity in quantum generative models with respect to simulatability and optimization}
\author[1]{Kaitlin Gili*}
\affil[1]{\small Department of Atomic and Laser Physics, University of Oxford, Oxford OX1 2JD \footnote{Both authors contributed equally to this work.}}
\affil[2]{\small University of Chicago, Chicago IL 60637}
\author[1,2]{Rohan S. Kumar *}
\author[1]{Mykolas Sveistrys}
\author[2]{Chris Ballance}
\maketitle

\begin{abstract}
Recent work has demonstrated the utility of introducing non-linearity through repeat-until-success (RUS) sub-routines into quantum circuits for generative modeling. As a follow-up to this work, we investigate two questions of relevance to the quantum algorithms and machine learning communities: Does introducing this form of non-linearity make the learning model classically simulatable due to the deferred measurement principle? And does introducing this form of non-linearity make the overall model's training more unstable? With respect to the first question, we demonstrate that the RUS sub-routines \emph{do not} allow us to trivially map this quantum model to a classical one, whereas a model without RUS sub-circuits containing mid-circuit measurements could be mapped to a classical Bayesian network due to the deferred measurement principle of quantum mechanics. This strongly suggests that the proposed form of non-linearity makes the model classically in-efficient to simulate. In the pursuit of the second question, we train larger models than previously shown on three different probability distributions, one continuous and two discrete, and compare the training performance across multiple random trials. We see that while the model is able to perform exceptionally well in some trials, the variance across trials with certain datasets quantifies its relatively poor training stability.
\end{abstract}

\section{Introduction}\label{s:intro}

In the field of Quantum Machine Learning (QML), the objective remains to gather evidence as to how and why quantum computation may be useful for learning tasks \cite{cerezo2022, Perdomo_Ortiz_2018, qml_advantage}. One of the most prominent research avenues is the design of quantum circuit architectures that can perform generative learning tasks - i.e. to train a model to learn an underlying probability distribution from an unlabeled training set such that it can subsequently generate new, high quality data \cite{gui2020review, ruthotto2021introduction, 2020Du, Rudolph2020GenerationOH, Zoufal2019, benedetti2019parameterized}. Currently, industry-scale classical generative models are deployed for applications in recommendations systems \cite{Recommender2016}, image restoration \cite{basioti2020image}, portfolio optimization \cite{alcazar2021enhancing}, and drug discovery \cite{GuacaMol2019}. Developing and characterizing more powerful generative models is of utmost importance for realizing the most advanced Artificial Intelligence applications \cite{gen_AI_apps}, and hence, the importance to investigate whether generative models designed with quantum mechanics leads to a computational complexity \cite{pac, hinsche2021learnability, hinsche2022} or alternative utility \cite{2019zhu,kgili_gen,alcazar2020classical} enhancement. 

Quantum circuits can represent probability distributions over a support of discrete bitstrings, and each quantum measurement in the computational basis is akin to generating a sample \cite{benedetti2019parameterized}. Thus, it is a natural idea to utilize these parameterized circuits as generative learners. We also have theoretical evidence that generative architectures with certain quantum gate structures have more expressive power than classical networks \cite{2020Du}. However, there are many questions to be answered before concluding that these models are useful, including: \emph{How can we efficiently train generally expressive quantum architectures?} \cite{abbas2023quantum} \emph{Can we intentionally design these architectures with inductive biases inspired by quantum mechanics that align with structures in specific datasets?} \cite{bowles2023contextuality, gili2023inductive, goyal2022inductive} \emph{Does designing quantum architectures with features from classical ML provide any utility?} \cite{cherrat2022quantum} 

Recent work proposed a quantum generative architecture that achieves non-linearity in the quantum state evolution using repeat-until-success (RUS) sub-routines containing mid-circuit measurements, similar to non-linear activations in a classical feed-forward network \cite{gili_qnbm}. This work put forth a preliminary investigation into the learning capabilities of the entire generative model, known as the Quantum Neuron Born Machine (QNBM), along with a demonstration of better performance over the more widely investigated Quantum Circuit Born Machine (QCBM) model \cite{Benedetti_2019, Liu_2018, 2020Coyle, gili_qcbm}, which does not contain these non-linear activations in the architecture. The work demonstrates evidence that the feature of non-linearity from classical ML might be useful in quantum circuit models. 

In this work, we attempt to answer two follow-up questions about the utility of this particular kind of non-linearity:\emph{(1) Does the non-linearity make the learning model efficiently classically simulatable as a result of the deferred measurement principle of quantum mechanics \cite{Nielsen_Chuang_2010}?} This principle states that an unconditioned qubits' measurement probability is unchanged no matter when the measurement is taken in the computation. This leads one to consider whether the mid-circuit measurements in the RUS sub-routine allow for the model to be simulated classically efficiently. A positive answer to this question would indicate that adding non-linearity in such a way reduces the model to one that only requires classical computation - an important insight for the field of quantum algorithms and ML. \emph{(2) Does the non-linearity lead to unstable optimization?} A positive answer to this question would indicate that perhaps the QNBM model is only useful in certain situations, where the initial parameters can be warm-started or chosen carefully and the data task aligns well. 

The work is organized as follows. In \secref{s:concept_overview}, we provide a general overview of generative modeling with quantum circuits and a detailed summary of the QNBM. Then, in \secref{s:results}, we provide evidence for the two questions above. Our results indicate that the mid-circuit measurements in the RUS sub-routines most likely \emph{do not} make the model classically simulatable with respect to the deferred measurement principle, but that they do increase the volatility of the model's optimization performance. We conclude with an outlook in \secref{s:outlook} that discusses future questions and investigations that are relevant for understanding non-linearity in quantum learning models. 

\section{Generative Models}\label{s:concept_overview}
The goal of an unsupervised generative model is to learn an unknown probability distribution $P_{target}$ from a finite set of data such that it can generate new samples from the same underlying distribution \cite{gui2020review, ruthotto2021introduction, zhao2018bias, gili2022evaluating}.  

Prior to demonstrating our results, we provide an overview of the model's circuit structure and training scheme. 
\subsection{Quantum Neuron Born Machine (QNBM)}\label{sec3:qnn}

\begin{figure}[h!]
\centering
\includegraphics[scale=0.3]{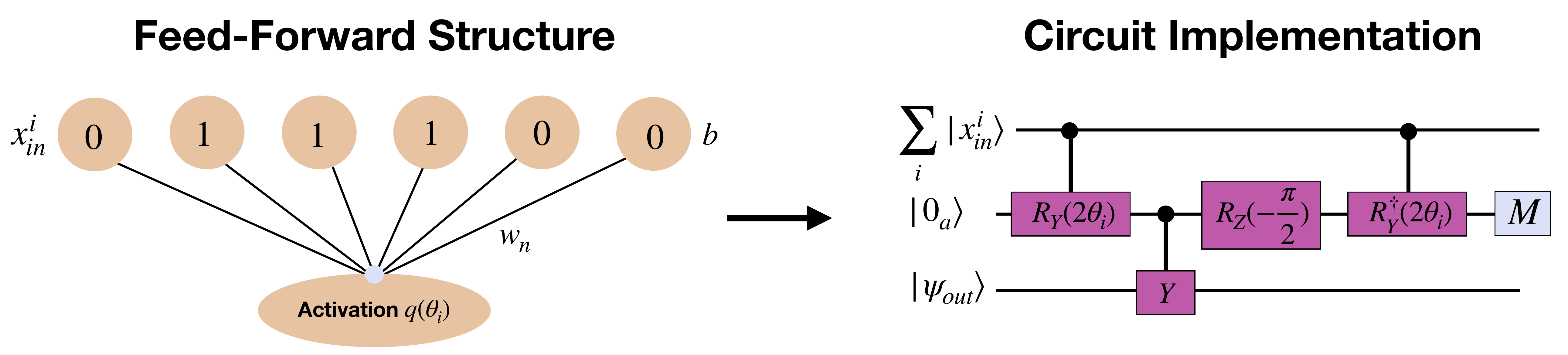}
\caption{\textbf{Visual demonstration of mapping information to a single output neuron in the QNBM.} On the left, we show the feed-forward structure of the neuron activation, which closely resembles a classical network containing trainable weights $w_{n}$ and biases $b$ on individual bitstrings $x^{i}_{in}$. The activation function $q$ introduces non-linearity to the output neuron in the next layer. To the right, we show the quantum circuit implementation that produces this non-linear mapping in the form of a quantum RUS circuit. The circuit feeds forward information from the superposition of bitstrings in the input layer $\sum_{i}|x_{in}\rangle$ and performs the non-linear activation on one output neuron $|\psi_{out}\rangle$. The QNBM is simply a multi-layered network comprised of these individual quantum neuron activations. Note that a main difference between the classical structure (left) and the quantum circuit model (right) is that the quantum network allows for a superposition of inputs in the initial layer. }
\label{fig:qneuron}
\end{figure}

The QNBM is a quantum analogue of a classical feed-forward neural network, introduced in Gili et al. \cite{gili_qnbm}. Each neuron in the network is assigned a qubit and is connected to the previous layer of neurons via a quantum neuron sub-routine \cite{cao2017quantum}, which is a Repeat-Until-Success (RUS) circuit \cite{Bocharov_2015, Paetznick_repeat, rus_proof}. The quantum neuron sub-routine is comprised of an input register $\ket{x_{in}}$ representing the previous layer of neurons, an output qubit initiated in the state $\ket{\psi_{out}}$ representing the neuron of the next layer, and also an ancilla qubit, initially in $\ket{0_{a}}$. The ancilla is used for mapping activation functions from the input layer of neurons $|x_{in}\rangle$ to each output neuron $\ket{\psi_{out}}$ in the next layer. A visual representation of the RUS sub-routine for a single neuron is demonstrated in \figref{fig:qneuron}.  

The RUS circuit performs a non-linear activation function at each output neuron after summing up the weights and biases from the neurons in the previous layer. Thus, each tunable parameter $\theta$ is a function of weights and biases: 
\begin{equation}
    \theta = w_1 x_1 + w_2 x_2 + ... + w_n x_n + b,
\end{equation}

\noindent
where ${w_n} \in (-1;1)$ are the weights for $n$ neurons in the previous layer and $b \in (-1;1)$ is the bias. Through mid-circuit measurements of the ancilla, the following activation function is enacted: 
\begin{equation}
q(\theta) = \arctan(\tan^2(\theta)).
\end{equation}

This non-linear activation function contains a sigmoid shape, making it comparable to those typically used in classical neural networks. By measuring the ancilla qubit to be $\ket{0_{a}}$, which occurs with a probability $p(\theta) > \frac{1}{2}$ \cite{cao2017quantum}, we perform the activation function on the output neuron. Otherwise, the activation function is not enacted and we must recover the pre-circuit state with an $X$ gate on the ancilla and $R_Y(-\pi/2)$ applied to the output qubit. The process will then be repeated until the ancilla measurement yields $\ket{0_a}$. 

The final state of each output neuron with a successful activation can be described as: 

\begin{equation}
\sum_{i} F_i|x^{i}_{in}\rangle \otimes |0_{a}\rangle \otimes R_{Y}q(2\theta_{i})|\psi_{out}\rangle, 
\end{equation}

\noindent
where $F_i$ is refers to an amplitude deformation in the input state during the RUS mapping and $\theta_{i}$ is the sum of weights and biases for each input bitstring. Note that due to the $R_{Y}$ rotation, the total function enacted on the output node is $\sin^2(q(2\theta_{i}))$. A key difference between the QNBM and classical neural networks is its ability to perform an activation function on a superposition of discrete bitstrings $\sum_i \ket{x^i_{in}}$.

QNBMs are defined by their neuron structure, i.e. the number of neurons in each layer \\
$(N_{in}, N_{hid}, ..., N_{out})$. At the end of the circuit, samples are drawn from the model according to the Born Rule when measuring \emph{only} the output neurons. These samples are used to approximate the model's encoded probability distribution, i.e. $P_{model}$.

The QNBM is trained with a classical optimizer to minimize the KL Divergence between $P_{target}$ and $P_{model}$  with a finite differences gradient estimator. Samples are generated post training for a separate evaluation of the model's ability to learn the desired distribution. This training scheme is very similar to other quantum generative models like QCBMs \cite{Liu_2018} and tensor networks \cite{han2018unsupervised}. 

\section{Results}\label{s:results}

In this section, we introduce a deeper investigation into the QNBM as a quantum generative model from multiple perspectives. First, we investigate how the neural network structure of the QNBM affects its classical simulatability, comparing the non-linear model to its "linearized" version. Next, we thoroughly assess the model's learning capabilities by training a $(5,0,6)$ network on three types of target probability distributions. To further distinguish the quantum model's capabilities from those of classical generative architectures, we benchmark the learning performance against a classical RBM, which contains a similar network structure to the QNBM with a similar number of resources. Lastly, we provide the first demonstration of the model's ability to generalize - i.e. learn an underlying ground truth probability distribution from a limited number of training samples. 

\subsection{\large\emph{Does the non-linearity make the learning model efficiently classically simulatable as a result of the deferred measurement principle?}}\label{sec1:express}

The deferred measurement principle states that an unconditioned qubit can be measured at any point in the computation and it's probabilistic outcome will not change \cite{Nielsen_Chuang_2010}. As the RUS non-linearity maps input qubits to the next layer of output qubits through a mid-circuit measurement protocol, it is possible to think that this mid-circuit measurement would provide one with information that can be used to simulate the probabilities classically. Here, we show that the RUS sub-routines \emph{do not} allow us to trivially map this quantum model to a classical one, whereas a model without RUS sub-circuits containing mid-circuit measurements could be mapped to a classical Bayesian network due to the deferred measurement principle of quantum mechanics.

Suppose we have a QNBM with $n$ layers, where the first $k < n$ layers are connected, and we want to connect a neuron in the $k + 1$'th layer to the network. The combined state of the first $k$ layers, the qubit in the $k+1$'th layer, and the ancilla can be generally written as
\begin{equation}
\ket{\psi_{k+1}} = \left(\sum^{2^{N_k}}_{i=1} \alpha_i \ket{\phi_i} \otimes \ket{x_i}_k\right) \otimes \ket{0}_{k+1} \otimes \ket{0}_a,
\label{qnbm_general_state}
\end{equation}
where each $\ket{\phi_i}$ describes the first $k-1$ layers, and each $\ket{x_i}$ is a single bitstring state representing the $k^{th}$ layer. Notice that if the network terminates at the $k^{th}$ layer, then the probabilities of the various output bitstrings $x_i$ are given by $|\alpha_i|^2$. By applying the gates as described in \secref{sec3:qnn}, we get the pre-measurement state
\begin{equation}
\begin{split}
    \ket{\psi_{k+1}} = \sum^{2^{N_k}}_{i=1} \alpha_i \ket{\phi_i} \otimes \ket{x_i}_k \otimes (\cos^2{\theta_i} \ket{0}_{k+1} \otimes \ket{0}_a \\ + \sin{\theta_i} \cos{\theta_i} \ket{0}_{k+1} 
 \otimes \ket{1}_a \\ + \sin{\theta_i} \cos{\theta_i} \ket{1}_{k+1} \otimes \ket{1}_a \\ + \sin^2{\theta_i} \ket{1}_{k+1} \otimes \ket{0}_a),
\end{split}
\end{equation}
where $\theta_i = (\sum^{2^{N_k}}_{j=1} w_{ij} x_j) + b_i$ .
If we measure the ancilla and obtain result zero, then up to normalization the state becomes
\begin{equation}
\begin{split}
    \ket{\psi_{k+1}} = \sum^{2^{N_k}}_{i=1} \alpha_i \ket{\phi_i} \otimes \ket{x_i}_k \otimes (\cos^2{\theta_i} \ket{0}_{k+1} \otimes \ket{0}_a \\ + \sin^2{\theta_i} \ket{1}_{k+1} \otimes \ket{0}_a).
\end{split}
\end{equation}
The probability of finding the connected neuron in the zero state and one state respectively become
\begin{equation}
P_0 = \frac{\sum^{2^{N_k}}_{i=1} |\alpha_i|^2 \cos^4{\theta_i}}{\sum^{2^{N_k}}_{i=1} |\alpha_i|^2 (\cos^4{\theta_i} + \sin^4{\theta_i})},
\end{equation}
\begin{equation}
P_1 = \frac{\sum^{2^{N_k}}_{i=1} |\alpha_i|^2 \cos^4{\theta_i}}{\sum^{2^{N_k}}_{i=1} |\alpha_i|^2 (\cos^4{\theta_i} + \sin^4{\theta_i})}.
\end{equation}
Let us compare these expressions to those from a "linearized" QNBM, where quantum neuron sub-routines are simply replaced with unitary Pauli-Y rotations. If we connect a neuron in the $k+1^{th}$ layer to the $k^{th}$ layer of neurons (starting from the same state as before), we obtain
\begin{equation}
\begin{split}
    \ket{\psi_{k+1}}^{(lin)} = \sum^{2^{N_k}}_{i=1} \alpha_i \ket{\phi_i} \otimes \ket{x_i}_k \otimes (\cos{\theta_i} \ket{0}_{k+1} + \\ \sin{\theta_i} \ket{1}_{k+1}),
\end{split}
\end{equation}
such that
\begin{equation}
P_0^{(lin)} = \frac{\sum^{2^{N_k}}_{i=1} |\alpha_i|^2 \cos^2{\theta_i}}{\sum^{2^{N_k}}_{i=1} |\alpha_i|^2 (\cos^2{\theta_i} + \sin^2{\theta_i})} = \sum^{2^{N_k}}_{i=1} |\alpha_i|^2 \cos^2{\theta_i},
\end{equation}
\begin{equation}
P_1^{(lin)} = \frac{\sum^{2^{N_k}}_{i=1} |\alpha_i|^2 \sin^2{\theta_i}}{\sum^{2^{N_k}}_{i=1} |\alpha_i|^2 (\cos^2{\theta_i} + \sin^2{\theta_i})} = \sum^{2^{N_k}}_{i=1} |\alpha_i|^2 \sin^2{\theta_i}.
\end{equation}
Notice that the probabilities are \textit{linear} combinations of the values $|\alpha_i|^2$ which also correspond to output bitstring probabilities if the network was terminated at layer $k$. Then the factors of $\cos^2{\theta_i}$ and $\sin^2{\theta_i}$ correspond to conditional probabilities. In other words, 
\begin{equation}
P_{k+1}^j = \sum^{2^{N_k}}_{i=1} P(j|i) P(i)
\end{equation}
where $P(j|i)$ means "probability to find a neuron in the $k+1^{th}$ layer in state $j$, given that the neurons in the $k^{th}$ layer have been found in state $i$. The probabilities are $P(i) = |\alpha_i|^2$, $P(j|i) = \cos^2(\theta_i)$ for $j = 0$ and $P(j|i) = \sin^2(\theta_i)$ for $j = 1$.

Since we are able to write the probabilities $P^{(lin)}_i$ in this form, sampling bitstrings from the $k+1^{th}$ layer of the network is equivalent to the process of sampling bitstrings from the $k^{th}$ layer, \emph{classically} calculating the probabilities for the $k+1^{th}$ layer, and then sampling it from those. The calculation of probabilities and subsequent sampling requires $O(E_{k \rightarrow k+1})$ calculations, where $E_{k \rightarrow k+1}$ is the number of edges connecting layers $k$ and $k+1$; therefore, such sampling is efficient. This is true for any pair of previously connected layers (\eqref{qnbm_general_state} describes a completely general QNBM state), and as such, one can sample efficiently and classically from the whole network with $O(E)$ calculations, where $E$ is the number of edges in the whole network. 

In fact, this construction is precisely a Bayesian network \cite{bayes_work}. Bayesian networks can be sampled from efficiently, and therefore the "linearized" QNBM can be efficiently sampled from as well. 

Clearly the same construction does not work for the non-linear QNBM as the probabilities for the connected neuron are not linear combinations of the output probabilities for the layer that precedes it, and so there are no well-defined "conditional probabilities". While of course there may be other classical models equivalent to the QNBM, we believe the connection is certainly very non-trivial and not due to the principle of deferred measurement. 

\subsection{\large\emph{Does the non-linearity lead to unstable optimization?}}\label{sec1:express}

\subsubsection*{Experiment Setup}

In this section, we challenge the QNBM to effectively learn three target probability distributions typically utilized in the literature for benchmarking: Bars and Stripes (BAS), cardinality-constrained, and discrete Gaussian. In doing so, our goal is to evaluate the non-linearity present in the QNBM by investigating its training effectiveness on these simple distributions. We significantly extend the investigation in \cite{gili_qnbm} by evaluating the model on three distributions with larger support than the original work. Furthermore, we include both discrete and continuous distributions and evaluate the model's training effectiveness on both types of distribution.

Each distribution is defined on $\{0,1\}^N$ , i.e. bit strings of length $N$. All simulations in our work are conducted with distributions of dimension $N=6$, as this constitutes the size of each network's output layer. A BAS distribution is composed of uniform probabilities over bit strings that represent either a bar or a stripe in a 2D binary grid. In this encoding, each binary digit represents a white $(0)$ or black square $(1)$ such that these patterns emerge \cite{Benedetti_2019}. For an $N=6$ distribution with a $2 \times 3$ grid, there are $20$ patterns for the model to learn \cite{Liu_2018}. A cardinality-constrained distribution contains uniform probabilities over bit strings that fit a given numerical constraint in the number of binary digits equivalent to $1$ (e.g. $001011$ has a cardinality of $3$). For a cardinality of $c = 3$, we have ${6 \choose 3} = 20$ patterns for the model to learn. Lastly, for the discrete Gaussian, we simply ask the model to learn the function $f: \{0,1\}^n \to [0,1]$,  $f(x) =\frac{1}{\sigma\sqrt{2\Pi}}\exp(-\frac{1}{2}\frac{(x-\mu)^{2}}{\sigma^{2}})$, where $\mu$ is the chosen mean and $\sigma$ is the chosen standard deviation. For all simulations, the distributions are peaked at the central bit string with $\sigma = 7$, as these values were appropriate for the distribution support size. 

In this part of the investigation, we provide the model with complete access to the underlying probability distribution, rather than using a finite number of training samples. This method, previously used in the literature to investigate alternative generative architectures such as Quantum Circuit Born Machines (QCBMs) \cite{2019zhu, Benedetti_2019, Liu_2018}, allows us to investigate two important attributes regarding the model's ability to learn a wider range of distributions: the model's expressivity and its ability to be effectively trained. 

The quality of the model with given weights can be evaluated by computing the overlap of the model's output probability distribution $P_{model}(x)$ with that of the target distribution $P_{target}(x)$. While there are many metrics that fulfil this purpose, it suffices for our small scale models to use the simple Kullback-Leibler Divergence \cite{kl} defined as: 

\begin{equation}
KL = \sum_x P_{target}(x) \log \left(\frac{P_{target}(x)}{\max(P_{model}(x), \epsilon)}\right), 
\end{equation}

\noindent
where $\epsilon << 1$ (in our case, $\epsilon \approx 10^{-16}$) such that the function remains defined for $P_{model}(x) = 0$. When assessing expressivity, a theoretically optimal model would obtain the following:

\begin{equation}
KL(P_{target}(x), P_{model}(x)) = 0 
\end{equation}

\noindent
This means that the model's distribution is identical to the target, and thus the model can fully express the target distribution. Thus, during training, the model uses the KL divergence as a loss function and searches for parameters that minimize this metric. Note that this evaluation provides no information about the model's capacity to generalize as the model receives all of the data from the underlying distribution as input. 

Here, we provide the distribution learning performance of a QNBM with small randomly initialized parameters on the three distributions. This QNBM has a $(N_{in}, N_{hid}, N_{out}) = (5,0,6)$ architecture, as it has been shown in previous work that QNBMs at small scales perform optimally when no hidden layers on introduced \cite{gili_qnbm}. With this constraint, we choose meta-parameters that enable each model to train optimally. The QNBM is trained with $2,000$ iterations and $10,000$ shots per iteration, for a total of 20 million shots throughout the training. When simulating the QNBM, we utilize post-selection rather than implement mid-circuit measurements with classical control for each RUS sub-routine. The number of shots quoted is the number of shots before post-selection. In practice, a quantum device running the QNBM would require the capacity for mid-circuit measurements and classical control to avoid an exponential shot cost. 

\subsubsection*{Experiment Results}

We simulate the performance of the QNBM for each distribution across $5$ independent trainings. The performance of each of these trainings is indicated by the final KL value achieved by the trained model. We assess the \textit{stability} of each training by inspecting the standard deviation of the KL values achieved by the model on these trainings, as well as the best, average, and worst trainings. These results are summarized in Table \ref{mem_results}.

\begin{table}[h]
\centering
\begin{tabular}{|l|r|r|r|l|}
\hline
                  & \multicolumn{1}{l|}{Minimum KL} & \multicolumn{1}{l|}{Maximum KL} & \multicolumn{1}{l|}{Average KL} & Standard Deviation          \\ \hline
Cardinality       & 0.0399                          & 0.0452                          & 0.0424                          & \multicolumn{1}{r|}{0.0024} \\ \hline
Bars and Stripes  & 0.0991                          & 0.2463                          & 0.1588                          & \multicolumn{1}{r|}{0.0786}                      \\ \hline
Discrete Gaussian & 0.0194                          & 0.4601                          & 0.2158                          & \multicolumn{1}{r|}{0.2009} \\ \hline
\end{tabular}
\caption{Accuracy Training Results.}
\label{mem_results}
\end{table}

\noindent
We observe a noticeable difference in stability when training the QNBM on each of the three distributions. In particular, training on the Cardinality distribution yields both a favorable minimum KL value and low standard deviation across independent training seeds. When training on the Bars and Stripes distribution, the model achieves moderate best performance and standard deviation. The training on the Discrete Gaussian distribution is particularly notable. In the best case, the model achieves remarkably low KL. However, in the majority of cases, the model's training is highly unstable, leading to a high final KL. The standard deviation of the model's trainings on this distribution is over 80x greater than that on the Cardinality distribution. 

\begin{figure}[H]
\centering
    \begin{subfigure}[b]{0.31\textwidth}
        \includegraphics[width=\textwidth]{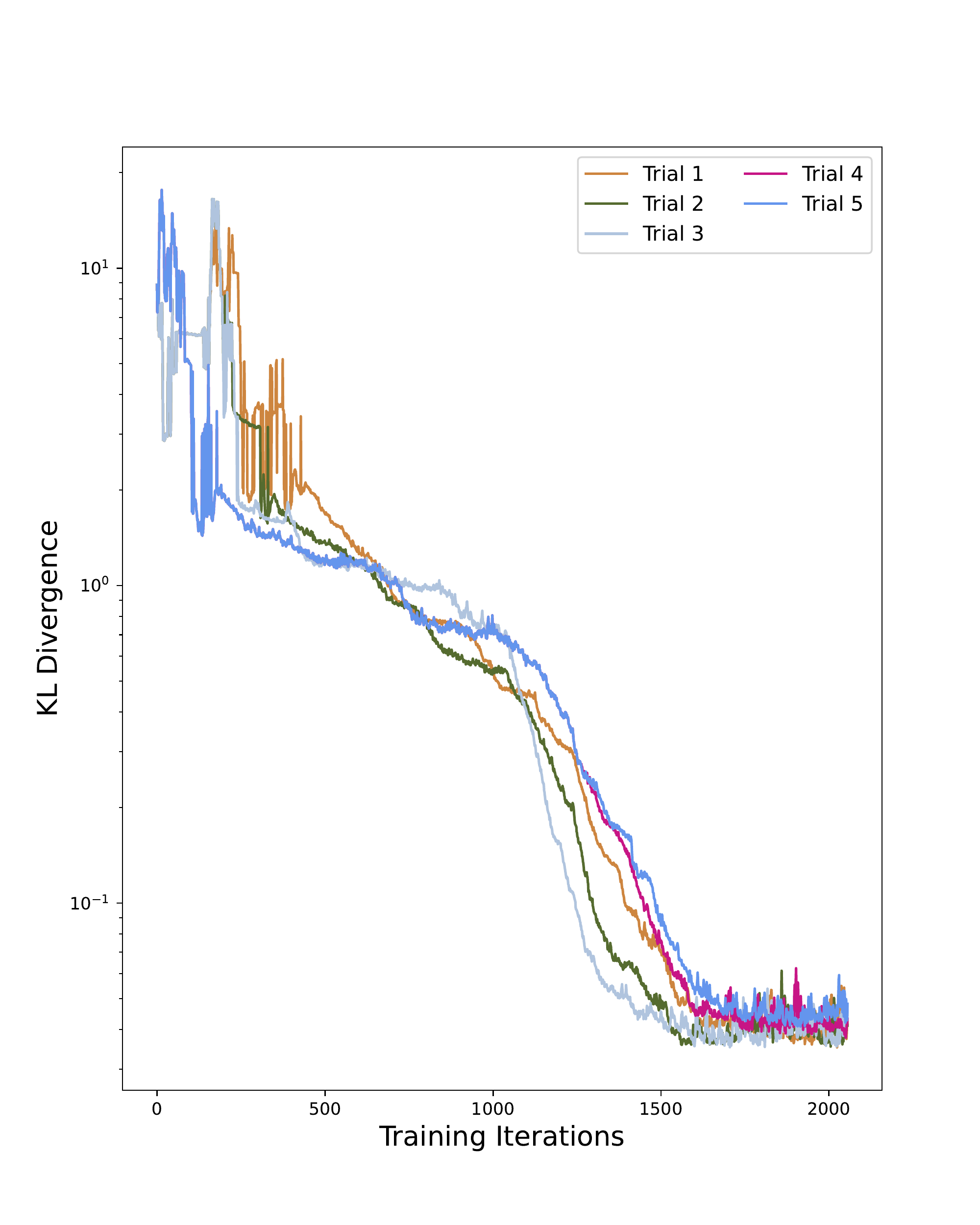}
        \caption{Cardinality Constrained}
        \label{sfig:qnbm_train_cardinality}
        \end{subfigure}   
    \begin{subfigure}[b]{0.31\textwidth}
        \includegraphics[width=\textwidth]{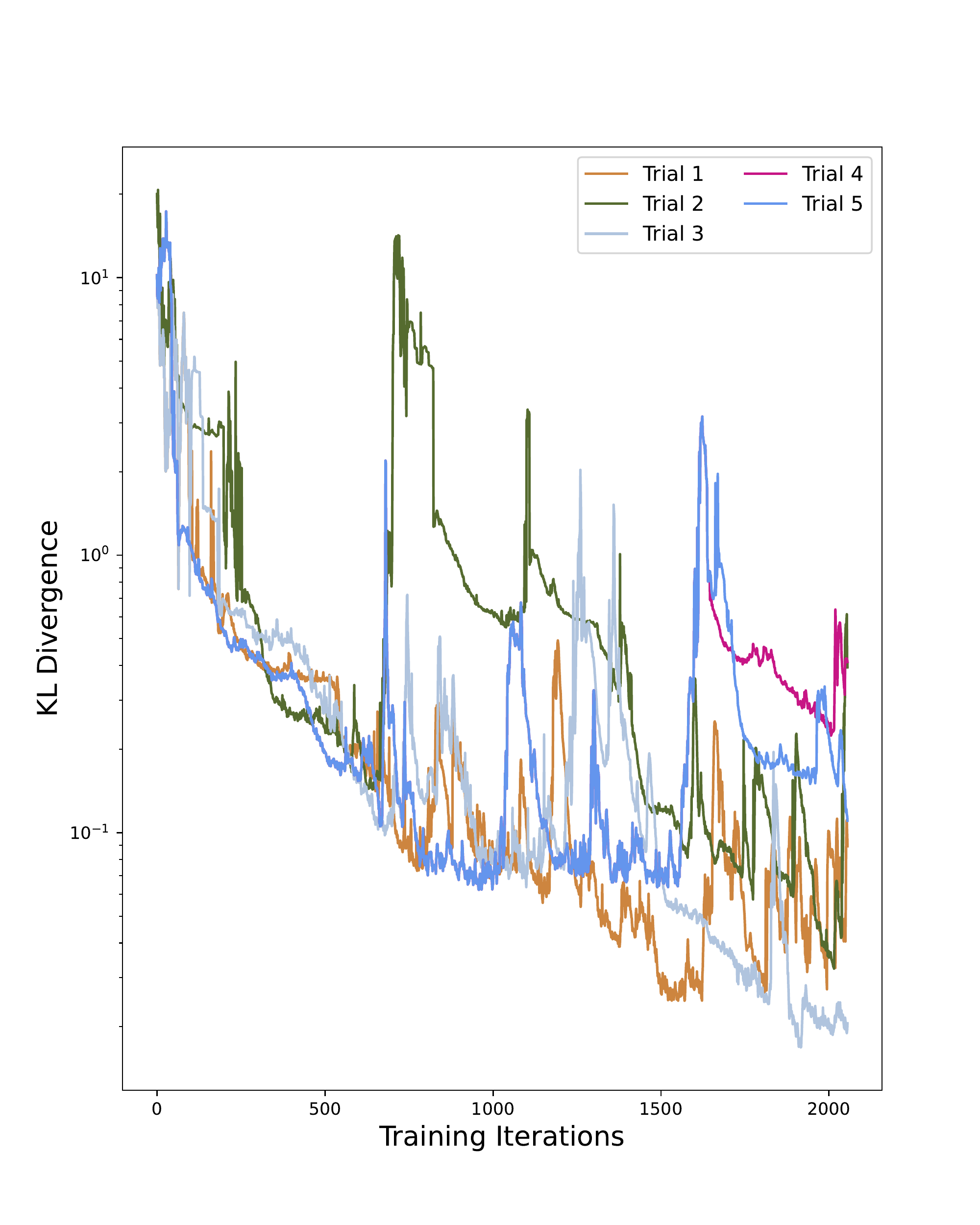}
        \caption{Discrete Gaussian}
        \label{sfig:qnbm_train_gaus}
        \end{subfigure}  
    \begin{subfigure}[b]{0.31\textwidth}
        \includegraphics[width=\textwidth]{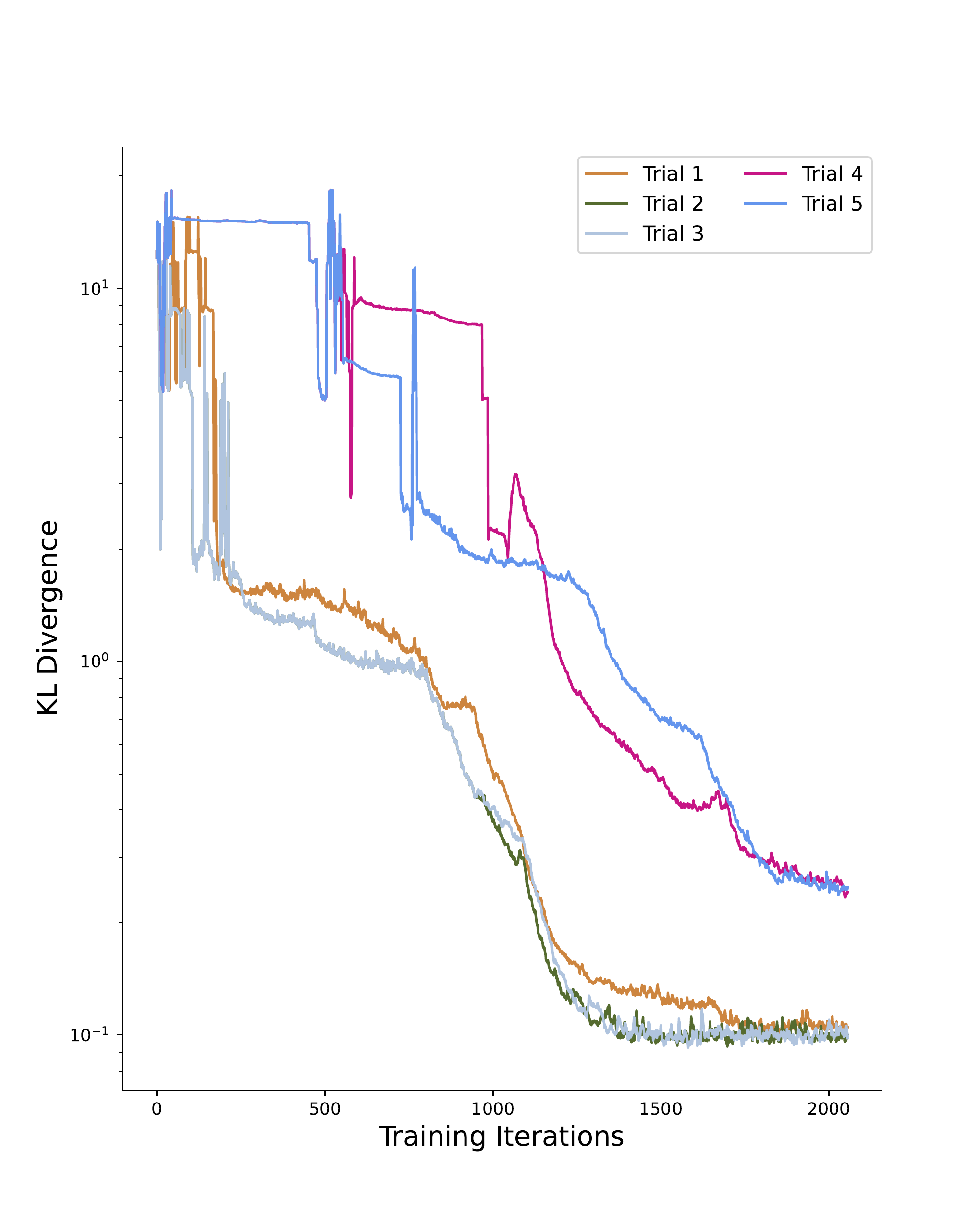}
        \caption{Bars and Stripes}
        \label{sfig:qnbm_train_bs}
        \end{subfigure}  
\captionsetup{justification=raggedright, singlelinecheck=false}
\caption{The KL Divergence vs. the number of training iterations for the QNBM across independent trials for three distributions.}
\label{sfig:qnbm_training_instability}
\end{figure}

\noindent
A visualization of the training performance and stability is shown in \figref{sfig:qnbm_training_instability}. As indicated by Table \ref{mem_results}, the QNBM's training on the Cardinality Constrained distribution is consistently stable, while its performance on the Bars and Stripes distribution is less so. The QNBM's training on the Discrete Gaussian distribution is markedly volatile, indicating an incompatibility between the model and this type of distribution. 


Overall, while the QNBM is able to achieve impressive learning performance as a standalone model in the best case, its training on even simple distributions can be volatile. 

\section{Outlook}\label{s:outlook}
In this work, we attempt to answer two follow-up questions to the insights revealed in Ref \cite{gili_qnbm}. First, we look more closely as to whether introducing non-linearity into quantum circuits for generative modeling through RUS sub-routines makes the model efficiently classically simulatable. The RUS sub-routines introduce non-linearity through mid-circuit measurements, which provide information regarding the qubits' state due to the deferred measurement principle, leading one to think these measurements could be used to formulate a classical Bayesian network. We demonstrate that this would hold in the case of taking mid-circuit measurements without RUS sub-routines, and does not hold for the RUS generated non-linearity. As such, the deferment measurement principle does not offer a trivial method to efficiently simulate the model with a classical computer.  

Secondly, we investigate whether the RUS sub-routines with mid-circuit measurements lead to training instabilities. As such, we train a $(5,0,6)$ network on three types of target probability distributions. We demonstrate that the model is expressive enough to capture the three types of probability distributions and can be effectively trained. However, the training is quite volatile over multiple random optimization trials, indicating that a good starting point is necessary with respect to a specific distribution to be learned. Perhaps this form of non-linearity is \emph{not as useful} for training general quantum neural networks and could be more useful for specific data tasks. 

We feel that this work provides further insight as to how non-linearity, a feature from classical ML, impacts quantum models. As such, some interesting questions remain for future investigation. While we now know that the mid-circuit measurement outcomes do not allow us to map the quantum model to a classical Bayesian one; we can investigate if those outcomes do provide us with any meaningful information that would help regularize the optimization. This is especially important if we see that non-linearity makes the training more unstable. Additionally, we could more theoretically investigate the inductive bias of the QNBM architecture and consider what datasets align well with its structure - i.e. what specific kinds of distributions is this model useful for? Is the non-linearity even necessary or does it only cause us more training pain at large scales? 

Because the model is so structurally close to that of a Bayesian network, we find it interesting to compare the two. Do the learning problems that we typically use Bayesian networks for also align well with these quantum networks? Do the quantum networks exhibit any advantages that are detectable? As Bayesian networks are useful for modeling cognition, are these models also useful for for this task \cite{bruza2015quantum, pothos2022quantum}? Does the non-linearity introduced into quantum circuit structures allow us to model states of cognition? 

Overall, we hope this work sheds more insight as to whether introducing non-linearity into quantum circuits for generative models is \emph{useful}. While it's more clear that the model requires a quantum device, it is also now less clear that the model will be stable across general learning tasks at larger scales. Further evidence is required to make larger claims around whether this feature of classical ML is important for quantum architectures.

\section{Acknowledgements}
The authors would like to recognize the Army Research Office (ARO) for providing funding through a QuaCGR PhD Fellowship. This work was supported by the U.S. Army Research Office (contract W911NF-20-1-0038) and UKRI (MR/S03238X/1). Additionally, the authors would like to recognize Marcello Benedetti for insightful conversations, especially in discussing classical simulatability, as well as feedback on the manuscript prior to submission. Lastly, the authors would like to acknowledge the places that inspired this work: Oxford, UK; Berlin, Germany; Ljbluiana, Slovenia; Bratislava, Slovakia; Naples, Soverato, Rimini, Rome, Italy; Istanbul, Capaddoccia, Turkey; Jerusalem, Israel; Singapore, Singapore; Lyon, Annecy, France; Geneva, Switzerland; Copenhagen, Denmark; Reykjavík, Akureyri, Egilsstadir, Vik, Selfoss, Ring Road, Iceland; Phuket, Thailand; and IL, CO, PA, MA, NJ, FL, NC, SC, USA. Thank you to all of the local people who became a part of the journey.

\printbibliography
\end{document}